\documentclass{aa}  
\usepackage{graphicx}
\usepackage{txfonts, psfig,subfigure,supertabular}
\begin{document}
   \title{A Multi-Epoch VLBI Survey of the Kinematics of CJF Sources \thanks{Figure 1 and Table 2 are only available in electronic form at the CDS via anonymous ftp to cdsarc.u-strasbg.fr (130.79.128.5) or via http://cdsweb.u-strasbg.fr/cgi-bin/qcat?J/A+A/}
			}

   \subtitle{Part I:  Model-Fit Parameters and Maps}

   \author{S. Britzen\inst{1,2,3} \and R.C. Vermeulen\inst{2}
   \and G.B. Taylor\inst{4,5}
   \and R.M. Campbell\inst{6}
   \and T.J. Pearson\inst{7} \and A.C.S. Readhead\inst{7} \and W. Xu\inst{7}
   \and I.W.A. Browne\inst{8} \and D.R. Henstock\inst{8} \and P. Wilkinson\inst{8}}

   \offprints{S. Britzen, sbritzen@mpifr-bonn.mpg.de}

   \institute{Max-Planck-Institut f\"ur Radioastronomie, Auf dem H\"ugel 69, D-53121 Bonn, Germany \and ASTRON, Netherlands Foundation for Research in Astronomy, Oude Hoogeveensedijk 4, P.O. Box 2, NL-7990 AA Dwingeloo, The Netherlands \and Landessternwarte, K\"onigstuhl, D-69117 Heidelberg, Germany \and Kavli Institute of particle Astrophysics and Cosmology, Menlo Park, CA 94025, USA \and National Radio Astronomy Observatory, Socorro, NM 87801, USA \and Joint Institute for VLBI in Europe, Oude Hoogeveensedijk 4, P.O. Box 2, NL-7991 PD Dwingeloo, The Netherlands \and California Institute
   of Technology, Department of Astronomy, 105-24, Pasadena, CA 91125, USA
   \and University of Manchester, Nuffield Radio Astronomy Laboratories, Jodrell Bank, Macclesfield, Cheshire SK11 9 DL, England UK}

   \date{Received ; accepted }

 
  \abstract
   {This is the first of a series of papers presenting VLBI observations of the 293 Caltech-Jodrell Bank Flat-Spectrum (hereafter CJF) sources and their analysis. 
   }
   {One of the major goals of the CJF is to make a statistical study of the apparent velocities of the sources. }
   {We have conducted global VLBI and VLBA observations at 5 GHz since 1990, accumulating thirteen separate observing campaigns. 
   }
   {We present here an overview of the observations, give details of the data reduction and present the source parameters resulting from a model-fitting procedure. For every source at every observing epoch, an image is shown, built up by restoring the model-fitted components, convolved with the clean beam, into the residual image, which was made by Fourier transforming the visibility data after first subtracting the model-fitted components in the uv-plane. Overplotted we show symbols to represent the model components.}
   {We have produced VLBI images of all but 5 of the 293
   sources in the complete CJF sample at several epochs and investigated the kinematics of 266 AGN.}
   \maketitle

\section{Introduction}
With a sample as large as the CJF, jet astrophysics (Lorentz factors, acceleration, propagation) is clearly addressed in important ways, through studying both morphologies and velocities. The completed CJF allows an investigation of the dependence of jet properties on a range of source parameters including host object type, luminosity, and redshift. 
Extensive VLBI surveys in the past have provided a morphological classification of compact radio sources (e.g., Wilkinson 1995 and references therein) and motion studies have yielded apparent velocity and Lorentz factor statistics that can be compared to other indicators of relativistic motion (e.g., Ghisellini et al. 1993; Vermeulen \& Cohen 1994, Vermeulen 1995; Jorstad et al. 2001; Kellermann et al. 2004; Cohen et al. 2006). 
Several projects have been conducted to investigate the pc-scale structures of samples of AGN using the VLBA, e.g., by Fomalont et al. (2000), Jorstad et al. (2001), Kellermann et al. (2004), and Piner et al. (2004). \\
The CJF survey integrates several VLBI surveys conducted from Caltech and Jodrell Bank into a complete flux-density limited sample. For all the CJ surveys, sources were selected from the region of sky bounded by declination $>$ 35$^{\circ}$ and Galactic latitude $|b|>10^{\circ}$ based on the 6~cm MPI-NRAO 5 GHz surveys (e.g., K\"uhr et al., 1981). The original sample (``Pearson-Readhead'', PR sample, Pearson \& Readhead 1981), is a complete sample of 65 sources with flux density $S_{5 \rm GHz}\geq$1.3 Jy, many of which were imaged with VLBI at 5 GHz and 1.6 GHz (Pearson \& Readhead 1981, 1988; Polatidis et al. 1995). The CJ1 extended the PR sample down to $S_{5 \rm GHz}\geq$ 0.7 Jy (total of 200 sources) (Polatidis et al. 1995; Thakkar et al. 1995; Xu et al. 1995). For CJ2 the limit is $S_{5 \rm GHz}\geq$ 0.35~Jy with the restriction that the sources should have a flat spectrum ($\alpha>-0.5$) resulting in a total of 193 sources (Taylor et al. 1994; Henstock et al. 1995). \\
The CJF, defined by Taylor et al. (1996), is
a complete flux-limited VLBI sample of 293 flat-spectrum radio sources, drawn
from the 6 cm and 20 cm Green Bank Surveys (Gregory \& Condon 1991; White \& Becker 1992)
with selection criteria as follows: $S$(6~cm)$\ge$ $350$ mJy,
{$\alpha _{20}^{6}$}{$\ge $}{$-0.5$}, $\delta $(B1950.0)$\ge $35$^{\circ }$, and {$|b^{{\rm II}}|$}{$\ge $}10$^{\circ}$. Although the CJF survey is based on surveys made at different epochs it can be regarded as a statistically complete survey (see Taylor et al. 1996, K\"uhr et al. 1981b).\\
Optical identifications have been made for 97 \% of the CJF sample and redshifts obtained for 94 \% of the objects (Stickel \& K\"uhr 1994; Stickel et al. 1994; Xu et al. 1994; Vermeulen \& Taylor 1995; Vermeulen et al. 1996, V\'{e}ron-Cetty \& V\'{e}ron 2003).
All redshifts now available for CJF are tabulated in Britzen et al. 2007a, with comments for values not previously published.\\
The composition of the CJF is 67 \% quasars, 18 \% radio galaxies, 11 \% BL Lac objects, and 3 \% still unclassified objects. Between 5 \% and 10 \% of the sources (Pearson et al. 1998) in the CJ samples are {\it compact symmetric objects} (Wilkinson 1995).
An overview summarizing
existing investigations of first- and second-epoch observations of subsamples of the CJF with
references is given in Pearson et al. (1998). Preliminary results for selected samples of CJF sources have already been published in Vermeulen (1995), Britzen et al. (1999, 2001) and Britzen (2002). A statistical analysis of 5 GHz VLBI polarimetry data from 177 sources in the CJF survey has been presented by Pollack et al. (2003).\\
The results presented here, serve as the basis for the analysis of the kinematics of the sources which will be discussed in Britzen et al. 2007a (hereafter Paper II). In Britzen et al. 2007b (hereafter Paper III) we present a correlation analysis between soft X-ray and VLBI properties of the CJF sources.\\
The CJF was designed as a state-of-the-art survey providing multiple
epochs of VLBI observations of a large, complete sample. To date, it is
indeed the largest multi-epoch survey of motions in terms of the number
of sources and jet components tracked. In addition, it provides angular resolution and dynamic range appropriate to identify and trace individual jet components reliably across the epochs. Due to its completeness, statistical statements can be made concerning the distributions of velocities, bending, pattern motions, and changes in the brightness of jet components and their dependence on the core separation.  \\
Given its size, completeness, and the range of source properties
spanned, this database should be of great utility for statistical
studies. Care was taken to ensure homogeneity in the observing strategy,
data reduction, and data analysis. We hope to have produced a body of
data that can be used to develop and test physical theories of active
nuclei in ways that have not previously been possible.
\section{Observations, data reduction and analysis}
\subsection{The observations}
Continued VLBI observations of the CJF sources have been performed since 1990 (see Table \ref{obs.tab}). Subsamples were observed in several global VLBI observations and in VLBA snapshot runs at 6 cm wavelength between March 1990 and December 2000.
The VLBA snapshot runs of CJF sources started in 1998. The observational strategy was to observe the sources 8 times in 5.5 minute snapshots in each observing session and to record the data over 32 MHz total bandwidths broken up into four baseband channels, with 1 bit sampling. The data were correlated in Socorro.\\ 
We aimed at a minimum of three epochs for every source since we found from experience that the unambiguous determination of the jet component position and motion requires at least
three observing epochs spread over roughly four years (minimum time span is less than one year). These observations are now complete; the last epoch for a subsample
of 34 sources was obtained in December 2000. In Table \ref{obs.tab} we list the correlator codes, dates, bandwidth, polarization information, antenna arrays, correlator, number and length of scans, and a reference for further information. 
\begin{table*}[htb]
\tabcolsep0.5mm
\hspace*{-3.5cm}
      \caption{Details of the global VLBI and VLBA observations.}
               \label{obs.tab}
\begin{minipage}{\textwidth}
{\scriptsize
\hspace*{-1.1cm}
         \begin{tabular}{lllllllll}
            \hline
            \noalign{\smallskip}
            Correlator Code  &  Date&Bandw.&Pol.&Antennas\footnotemark&Correlator& Scans&Ref. \\
            \noalign{\smallskip}
            \hline
            \noalign{\smallskip}
            R51     &March 1990& 2 MHz&single&SBWLNoKGFLaPtKpYO&JPL/Caltech Block II corr. &3x20-30 min.&Xu et al. 1995\\
            GX2F    &September 1991  &2 MHz&single&SBWJ2LNoCKnTDeDaKGNlFdLaPtKpBrYO&JPL/Caltech Block II corr.&3x20-30 min.&Xu et al. 1995\\
            GX2G    &March 1992 & 2 MHz&single&SWJ2LNoCKGNlFdLaKpOvBrY&JPL/Caltech Block II corr.&3x20-30 min.&Xu et al. 1995\\
         GV10/GW5&05.06.--07.06.1992& 2 MHz&single&SBWLNoKGNlFdLaPtKpOvBrY&JPL/Caltech Block II corr.&3x20-30 min.&Taylor et al. 1994, Xu et al. 1995\\
         GW7     &24.09.--27.09.1992& 2 MHz&single&SBWLNoKGNlFdLaPtKpOvBrY&JPL/Caltech Block II corr.&3x20-30 min.&Taylor et al. 1994, Xu et al. 1995\\
           GW7b  &01.03.--02.03.1993& 2 MHz&single&J2BSWLNGYBrHnKpNlOvPtSc&Caltech Block II corr.&3-4x20 min.&Taylor et al. 1994\\
           GW10     & 09.06.1993--16.06.1993 &2 MHz &single&J2 BSWLNUYBrHnNlOvPt&Caltech Block II corr.&3-4x20 min.&Henstock et al. 1995\\
            BV15A& 25.08.1995 1200 UT -- 29.08.1995 1200 UT&8 MHz &single&VLBA&Socorro&8x6.5 min.&Taylor et al. 1996\\
            BV15B & 03.09.1995 1200 UT -- 05.09.1995 1200 UT&8 MHz &single&VLBA&Socorro&8x6.5 min.&Taylor et al. 1996\\
            BV019   & 17.08.1996 1200 UT -- 25.08.1996 2200 UT &8 MHz &dual&VLBA&Socorro&8x6.5 min. &\\
            BV025A  & 08.02.1998 0900 UT -- 09.02.1998 0900 UT &32 MHz&dual&VLBA&Socorro&8x5.5 min. &  \\
            BV025B  & 12.02.1998 0300 UT -- 14.02.1998 0300 UT &32 MHz&dual&VLBA&Socorro&8x5.5 min. &\\
            BV025C  & 20.02.1998 0600 UT -- 22.02.1998 0600 UT & 32 MHz&dual&VLBA&Socorro&8x5.5 min.&\\
            BB119A  & 21.11.1999 1200 UT -- 22.11.1999 1200 UT & 32 MHz&dual&VLBA&Socorro&8x5.5 min.&\\
            BB119B  & 23.11.1999 2000 UT -- 24.11.1999 2000 UT & 32 MHz&dual&VLBA&Socorro&8x5.5 min.&\\
            BB119C  & 26.11.1999 0700 UT -- 27.11.1999 0700 UT & 32 MHz&dual&VLBA&Socorro&8x5.5 min.&\\
            BB131   & 16.12.2000 0200 UT -- 17.12.2000 1200 UT &  32 MHz&dual&VLBA&Socorro&8x5.5 min.&\\
            \noalign{\smallskip}
	    \hline
         \end{tabular}
	 }
	 \noindent
\footnotetext{$^{1}$The antenna letter codes (telescope names and locations in brackets) stand for: S (Onsala, Sweden), B (Effelsberg, Germany), W (WSRT, Netherlands), J2 (JBNK, Jodrell Bank, UK), U (Simeiz, Crimea, Ukraine), L (Medicina, Italy), No (Noto, Noto, Italy), C (Cambridge, Cambridge, UK), Kn (Knockin, Knockin, UK), T (Tabley, Tabley, UK), De (Defford, Defford, UK), Da (Darnhall, Darnhall, UK), K (Haystack, Westford, MA, USA), G (NRAO, Green Bank, WV, USA), F (FDVS, Fort Davis, TX, USA), Hn (VLBA\_Hn, Hancock, NH, USA), Nl (VLBA\_Nl, North Liberty, IA, USA), Fd (VLBA\_Fd, Fort Davis, TX, USA), La (VLBA\_La, Los Alamos, NM, USA), Sc (VLBA\_Sc, Saint Croix, VI, USA), Pt (VLBA\_Pt, Pie Town, NM, USA), Kp (VLBI\_Kp, Kitt Peak, AZ, USA), Ov (VLBA\_Ov, Owens Valley, Ca. USA), Br (VLBA\_Br, Brewster, WA, USA), Y (VLA, Socorro, NM, USA), O (OVRO, Owens Valley, CA, USA).}
\end{minipage}
 \end{table*}
\subsection{The data reduction}
To create a homogeneous, statistically valid database, we started
a systematic (re-)analysis of all epochs for all sources obtained in the 1990s.
All sources and epochs of all ``old'' (data before 1998) and ``newly'' obtained data sets have been analyzed in the same standardized way. Despite using a global array for the older epochs and the VLBA for the new epochs, we aimed at obtaining similar observing conditions, calibration methods (to ensure a reliable calibration of the sources, we included in each observing run at least one calibrator source, 3C 279, which was observed at similar $(u, v)$ ranges), and reduction techniques. Calibration and fringe-fitting were done using standard procedures in the Astronomical Image Processing System (AIPS, Greisen 1990). With automated mapping within {\it difmap} (v.2.4b, Shepherd 1997), making use of the script {\it automap}, we obtained Clean maps of similar quality for all data sets.
Extended sources were reprocessed using {\it shift} (within {\it difmap}), to move the observation phase-center. In addition, we re-{\it clean}ed with a different pixel size to map the extended emission reliably as well.
A critical analysis of the VLBI jet components and a comparison of jet properties requires a quantitative determination of the jet components' features. We therefore fitted Gaussian model components directly to the observed visibilities (real and imaginary parts) using the Levenberg-Marquardt non-linear least squares minimization technique (program {\it modelfit} within {\it difmap}) to fit the brightness, sizes and positions of the individual jet components.\\
We modelfitted all sources at all epochs independently, starting from a
point source, and using circular Gaussian components, with parameters
flux density ($S$), position in Cartesian coordinates ($x,y$), and
semi-diameter ($M$). The positions were later also converted to polar
coordinates ($r,\theta$), as explained below. We used circular components since it turned out that the estimations of the axial ratio parameters were often enough ill-conditioned to make it unjustifiable to include them as free parameters in the model (i.e., they are very highly correlated with other parameters).
The sizes of the circular jet components were allowed to vary between epochs. We stopped adding jet components within the model-fitting process whenever a solution had been obtained, such that adding an additional component would not improve the quality of the fit, i.e. reduce the value for chi-square, significantly.\\
\subsection{The uncertainties}
We calculated (statistical) uncertainties for the fitted Gaussian parameters for each
source at each epoch via a slight modification to {\it difmap}.  We derived the covariance matrix, $\mathcal{C}$, from the Hessian matrix (which was already computed during the model-fitting procedure) by 
using the pre-existing {\it lm\_covar} function, which simply performed 
Gauss-Jordan elimination to invert the matrix.  The uncertainty for the $i^{\rm th}$
parameter derives from the square-root of the diagonal elements, $\sigma_i = \sqrt{\mathcal{C}_{ii}}$, 
and the correlation matrix can be constructed from $\mathcal{C}_{ij} / \sqrt{ |\mathcal{C}_{ii}\mathcal{C}_{jj}| }$.
These matrix operations were done while still in the ($S,x,y,$MAJ) parameterization for the circular Gaussian components.
We then shifted the reference point to $(x,y)=(0,0)$, and shifted the positions of all other 
components accordingly: $(x'_i,y'_i) = (x_i-x_r, y_i-y_r)$; $r$ and $\theta$ followed
from these:  $r_i=\sqrt{{x'_i}^2 + {y'_i}^2}$ and $\theta_i = \tan^{-1}({x'_i}/{y'_i})$. Table \ref{modelfit} lists $(x',y')$ and $(r,\theta)$, with the uncertainties scaled for a reduced chi-square
of unity (i.e., multiplied by $\sqrt{\tilde\chi^2}$).
We computed the uncertainties in $r$ and $\theta$ through
standard error-propagation formulae, including the correlations among the parameters 
involved ($x_i,y_i,x_r,y_r$).  Explicitly, these are:
\begin{eqnarray*}
   \sigma_{r_i} &= \frac{1}{r_i} &[\ {x'_i}^2 (\sigma_{x_r}^2 + \sigma_{x_i}^2 - 2\sigma_{x_ix_r}\sigma_{x_i}\sigma_{x_r}) \\
                &     & \mbox{} + {y'_i}^2 (\sigma_{y_r}^2 + \sigma_{y_i}^2 - 2\sigma_{y_iy_r}\sigma_{y_i}\sigma_{y_r}) \\
                &     & \mbox{} + 2x'_iy'_i (\sigma_{x_ry_r}\sigma_{x_r}\sigma_{y_r} - \sigma_{x_ry_i}\sigma_{x_r}\sigma_{y_i})\\
                &     & \mbox{} + 2x'_iy'_i (- \sigma_{x_iy_r}\sigma_{x_i}\sigma_{y_r} + \sigma_{x_iy_i}\sigma_{x_i}\sigma_{y_i}) ]^{1/2};\\
   \sigma_{\theta_i} &= \frac{1}{r_i^2} &[\ {x'_i}^2 (\sigma_{y_r}^2 + \sigma_{y_i}^2 - 2\sigma_{y_iy_r}\sigma_{y_i}\sigma_{y_r}) \\
                &     & \mbox{} + {y'_i}^2 (\sigma_{x_r}^2 + \sigma_{x_i}^2 - 2\sigma_{x_ix_r}\sigma_{x_i}\sigma_{x_r}) \\
                &     & \mbox{} + 2x'_iy'_i (\sigma_{x_ry_i}\sigma_{x_r}\sigma_{y_i} + \sigma_{x_iy_r}\sigma_{x_i}\sigma_{y_r})\\
                &     & \mbox{} + 2x'_iy'_i (- \sigma_{x_ry_r}\sigma_{x_r}\sigma_{y_r} - \sigma_{x_iy_i}\sigma_{x_i}\sigma_{y_i}) ]^{1/2}.\\
\end{eqnarray*}
Customary caveats associated with statistical uncertainties apply here:  array composition/sensitivity
and (u, v) coverage can vary across epochs for a given source.\\
An alternative program to calculate the uncertainties is provided by the program {\it difwrap} (Lovell 2000). 
For comparison, we calculated the uncertainties from difwrap for a sample of randomly selected sources (0018+729 first epoch, 0110+495 first epoch, 0800+618 third epoch, and 1629+495 second epoch). 
We compared the uncertainties of the flux-density, core separation, position angle, and size of major axis calculated by both programs and found that on average the {\it difwrap}-uncertainties are smaller by $\sim$ 20 \%. We thus believe the uncertainties calculated by our program represent conservative values. These uncertainties in and correlations among the estimated circular Gaussian parameters will serve as the {\it a priori} covariance matrices in the estimation of jet-component kinematics presented in Paper II.
\subsection{Component identification}
The identification of jet components across epochs was done through
extensive inspection of all images and models, looking for consistency
in the flux density evolution and the relative positions and movements
of features. Components can be stationary, or move outwards or inwards.
Components which were clearly identifiable in at least two epochs
received labels, using the naming convention discussed in \S\ref{results}. Not all
epochs show all components. In some sources, jet component
identification was problematic. Jet components can fade or be masked by
other brighter components. An individual jet component may appear to
split into two features from one epoch to another, and conversely two
features may appear to merge. When it was obvious to us that such
features were in fact part of a single entity, one component label was
given. In some cases we suspected that features which have rather
different flux densities, position angles, and/or core separations in
different epochs are nevertheless associated with each other, but when
the correspondence was too uncertain these features have not been
identified with any component labels. The issues of splitters and
mergers, and the choice of the reference feature, are obviously
particularly important for the motion analysis, and are further
discussed in Paper II, where we also assign a reliability ``quality
factor'' to each component which has a definable motion.
\subsection{Number of components per jet}
Based on those jet components that have been taken into account for the determination of the proper motions we find that the average galaxy-jet has 3.6 jet components (based on 47 sources), those of quasars 2.7 (based on 180 sources), and BL Lac Objects 2.9 jet components (31 sources). 8 still unclassified sources have not been considered here. On average, galaxies tend to have longer pc-scale radio-jets than quasars or BL Lac Objects.\\

\section{Results}
\label{results}
We cleaned all the images and performed model-fitting in the (u, v) plane to optimally reproduce the clean image. In Fig.~\ref{maps} we show these clean maps with symbols representing the Gaussian model components overplotted. We do not show 0218+357 (a gravitationally lensed source) and five other sources where imaging was problematic, as discussed in \S\ref{problems}.
The individual circular jet component positions and sizes are indicated in this figure by encircled crosses. In Table \ref{modelfit} we present the results of the model-fitting procedure. 
We list the CJF source name in B1950.0 and J2000 coordinates (JVAS: Wilkinson et al. 1998, Browne 1998; V\'{e}ron-Cetty \& V\'{e}ron 2001; CLASS: Myers et al. 2003; NVSS: Condon et al.1998), the epoch of observation 
(year/month/day), and for each component its identification,
flux density (S), position with respect to the reference point (both as $(x,y)$ and $(r,\theta)$), and the size of the major axis (M). All parameters have associated uncertainties, as described in \S2.3. In addition, we list the reduced chi-squared value and in brackets the number of degrees of freedom which shall serve as a measure for the quality of the fit.\\
In the component identification, the letter $r$ denotes the reference point (in most cases this is the brightest component in the source --- the ``core''), the letter C (followed by a number, increasing with increasing separation from the core) denotes 
a component of the main jet, and CC (plus a number) denotes a component on the counter-jet side. A blank identification means that the component could not be detected confidently in any of the other epochs. The components are sorted according to increasing separation from the core.
In some cases, especially when the first jet component becomes brighter than the core, 
it is questionable where the core really is.  In these cases, we choose the reference point to be at ``one end'' of the jet.\\
A table listing details of the CJF-sample sources (source names, coordinates, flux-densities, spectral indices) can be found in Taylor et al. 1996.
A table listing the source redshifts and optical identifications and all parameters relevant for the motion analysis will be given in Paper II.
\subsection{Problematic observations}
\label{problems}
The following sources are not included in Table \ref{modelfit} or any further analysis: 0256+424, 0344+405, 0424+670, 0945+664, 1545+497. 
These sources were either too faint during the time of the observations or wrong coordinates were used for observing (0344+405). Three sources (0824+355, 0954+556, 1642+690) were too faint in one epoch, leaving only two usable for tracing component motion.\\ 
\subsection{Point-like sources}                                                                         The following sources appear to be point-like in at least three of the epochs: 0621+446, 0636+680, 1254+571, 1308+471, 1342+663, 1417+385, 1638+398, 1851+488, 2005+642. Some sources appear to be point-like in one epoch: 0016+731, 0615+820, 1125+596, 1818+356, 1839+389.\\

\subsection{Complicated jet structures}
Some sources (e.g., 0604+728, 0627+532, 0950+748, 2255+416) have rather complicated jet structures. Our snapshot observations reveal the basic structures and jet components. To trace jet component motion in better detail, more frequent observations with higher dynamic range would be required.\\

\section{Comments on individual sources}
In the following, we comment on sources that reveal peculiarities. Where possible, we compare 
our results with the results presented in the literature. Many of the CJF sources were also observed in the VSOP pre-launch survey by Fomalont et al. (2000, hereafter F2000), based on 5 GHz VLBA observations in June 1996; the 2 cm survey by Kellermann et al. (1998, hereafter K98; Zensus et al. 2002), based on 15 GHz VLBA observations in 1994--2002; the VSOP AGN Survey (Hirabayashi et al. 2000, Scott et al. 2004); and/or the USNO sample investigated by Fey \& Charlot (1997, hereafter FC97) based on 2.32 GHz and 8.55 GHz VLBA 
observations from April 1995 and October 1995. Since these observations were performed around the same time as our observations, we favor these works for comparison.\\
\begin{flushleft}
{\it 0016+731}\\
\end{flushleft}
We identify and trace one component and find evidence for another (unlabeled) component at a significantly different position angle ($\Delta\theta \sim 120 ^{\circ}$). 
This difference in position angle for the different jet components is
supported by FC97, who obtained a similar difference for different components from observations at 2.32 GHz and 8.55 GHz. F2000 observe one jet component at $r=2.8$ mas, $\theta=180^{\circ}$, which is consistent with our second jet component in our second epoch. Additional support for significantly changing angles with higher resolution is provided by Scott et al. (2004) in VSOP observations performed at 5 GHz.\\
\begin{flushleft}
{\it 0035+367}\\
\end{flushleft}
The identification is complicated by the faintness of the jet components. We trace only one jet component (C1) in this source. The source is a likely candidate for fast motion with structure that is difficult to model. \\ 
\begin{flushleft}
{\it 0035+413}\\
\end{flushleft}
The results of F2000 are in excellent agreement with
our results from the first epoch: C1, $r$ = 1.14 mas {\it vs.\ }1.19 mas; C3, $r$ = 6.67 mas {\it vs.\ }6.73 mas; C4, $r$ = 12.28 mas {\it vs.\ }12.33 mas. \\
\begin{flushleft}
{\it 0102+480}\\
\end{flushleft}
Two components can be traced on opposite sides of the core. F2000 confirm C1 with a jet 
component at $\sim$ 0.5 mas, position angle --158$^{\circ}$. They, in addition, find a component at $\sim$ 0.8 mas and 30$^{\circ}$ (most likely CC1). 
This source is a GPS source according to Marecki et al. (1999).\\
\begin{flushleft}
{\it 0153+744}\\
\end{flushleft}
This is a well-known source which never revealed motion, despite having been observed 
several times. Hummel et al. (1997) confirm stationarity of component B and place 
an upper speed limit of (0.007$\pm$0.025) mas/year. We identify eight jet components within this complicated source and confirm the results of Ros et al. (2001).\\ 
\begin{flushleft}
{\it 0205+722}\\
\end{flushleft}
The model-fitting process yielded a component at $\sim$37 mas core separation. 
In the model-fits presented here, we ignore this component. According to Augusto et al. (1998), this
source reveals extended structure in VLA and MERLIN observations and
can be classified as a MSO (medium symmetric object) with a bright core. We label and trace the components but we do not consider this source in the motion analysis since the component identification is not sufficiently unambiguous. \\
\begin{flushleft}
{\it 0212+735}\\
\end{flushleft}
This source most likely shows backward motion of one bright jet component (C2). F2000 confirm three jet components at 1.8 mas (C2 in this paper), 6.4 mas (C4 in this paper), and 13.9 mas (C5 in this paper), at position angles $\sim$104$^\circ$, 104$^\circ$, and 92$^{\circ}$,
respectively. The brightest component labeled as `r' in this paper might not be identical to the true core in this source since the first jet component also reveals a high flux density and in epoch 3 this component is even brighter than `r'.\\
\begin{flushleft}
{\it 0218+357}\\
\end{flushleft}
This is a gravitational lens (e.g., Patnaik et al. 1993). The dynamic range and angular resolution of the observations presented here are different for different epochs. This complicates a possible identification (and labeling) of jet components through the epochs and we do not consider this source in our further investigations.  \\
\begin{flushleft}
{\it 0249+383}\\
\end{flushleft}
We can trace four jet components reliably based on the second and third epoch. However, we cannot identify components in the first epoch consistently with the other epochs, so we ignore the first epoch in the subsequent kinematic modeling. Whether the scenario presented here is valid could only be decided with the help of further observations.\\ 
\begin{flushleft}
{\it 0316+413}\\
\end{flushleft}
This source (3C 84, NGC 1275) reveals a complex and bright jet structure (see also e.g., Krichbaum et al. 1992; Walker et al., 1994; Dhawan et al. 1998; Homan \& Wardle 2004). Assuming that the position angle remains roughly the same across the epochs, we trace jet components that approach and others that separate from the core. Further observations are required to trace the jet component motions in more detail.  \\
\begin{flushleft}
{\it 0402+379}\\
\end{flushleft}
This is a CSO and binary black hole candidate (Maness et al. 2004). A complicated jet structure in this source aggravates any component identification. To facilitate the inter-epoch comparison, we chose a reference point that is not identical to the brightest component of the jet, but seems to be the most compact object of the jet. We do not consider this source in any further calculations. \\ 
\begin{flushleft}
{\it 0537+531}\\
\end{flushleft}
In our identification scenario the jet component closest to the core (in epoch 1 and 4 and unlabeled in the table) is part of the core, i.e., not resolved as component.\\
\begin{flushleft}
{\it 0546+726}\\
\end{flushleft}
This source reveals a clear double source morphology.\\
\begin{flushleft}
{\it 0600+442}\\
\end{flushleft}
The simplest identification scenario for this source suggests that three jet components approach the reference point. Further observations are needed to monitor jet component motion in this source and to confirm the direction of motion.\\
\begin{flushleft}
{\it 0604+728}\\
\end{flushleft}
The jet emission around the core (r) appears at changing position angles. Our observations do not allow a detailed analysis of this phenomenon. Higher-frequency data might reveal whether the ejection angle of jet component varies in this source.\\  
\begin{flushleft}
{\it 0615+820}\\
\end{flushleft}
No jet component can be reliably identified and traced in this source. \\
\begin{flushleft}
{\it 0636+680}\\
\end{flushleft}
This source reveals a point-like structure in our observations, as
supported by FC97. \\
\begin{flushleft}
{\it 0650+371}\\
\end{flushleft}
The flux-density changes in this source are drastic. We therefore do not include this source in our analysis of the kinematics.\\
\begin{flushleft}
{\it 0650+453}\\
\end{flushleft}
The position angles of the individual jet components are too different between the epochs and prevent any reliable jet component identification. In other words, any jet component identification would lead to large position angle changes between the epochs. We therefore do not include any features in this source in the kinematic analysis in Paper II.\\
\begin{flushleft}
{\it 0707+476}\\
\end{flushleft}
Choosing the proper reference point is difficult in this source. In addition, the chosen reference component (r) shows significantly different flux-densities in the four different epochs. Thus, variability --- supposing we labeled the same component in every epoch --- further complicates the identification of components. \\
\begin{flushleft}
{\it 0716+714}\\
\end{flushleft}
Although this IDV-source (Wagner \& Witzel 1995, and references therein) reveals a simple structure, significantly different identification scenarios for the jet components for this IDV source have been published in the literature (e.g., Bach et al. 2005, Jorstad et al. 2001, Britzen et al. 2006). Our identification scenario is supported by the assignment of Bach et al. (2005), where four of the five epochs presented in this paper here are incorporated into an identification scenario based on a larger number of observations obtained at different frequencies.\\
\begin{flushleft}
{\it 0718+793}\\
\end{flushleft}
We do not find any convincing identification scenario for jet components in this source.\\
\begin{flushleft}
{\it 0746+483}\\
\end{flushleft}
The jet straightens out in the last epoch, the wiggling is significantly more prominent in the first and second epoch.\\
\begin{flushleft}
{\it 0749+540}\\
\end{flushleft}
The jet component in the third epoch cannot reliably be identified with either of the two components from the other epochs.\\
\begin{flushleft}
{\it 0821+394}\\
\end{flushleft}
We do not include the jet components from the first epoch in our identification since no component can reliably be traced. We find another component at a core separation of 294.5 mas and position angle of --49.7$^{\circ}$ (not listed in the table). This component and position is exactly supported by F2000. \\
\begin{flushleft}
{\it 0821+621}\\
\end{flushleft}
Although clearly a second jet component is visible in the table, we do not label this component since the position in core separation would vary un-physically across the epochs.\\
\begin{flushleft}
{\it 0824+355}\\
\end{flushleft}
The (u, v) coverage of the first epoch is less dense than in the other two epochs. We thus ignore this epoch in this paper.\\
\begin{flushleft}
{\it 0831+557}\\
\end{flushleft}
For this source the position of the ``true'' core is unknown. If our identification of the core position is correct, then the core shows significant flux-density variability.\\
\begin{flushleft}
{\it 0942+468}\\
\end{flushleft}
The third epoch for this source reveals a significantly different position for component C2 than expected from the other epochs. We thus do not label this component in this epoch. Future monitoring of this source might clarify possible reasons for this outlier.\\
\begin{flushleft}
{\it 0954+556}\\
\end{flushleft}
The source was very faint in two of the observed four epochs. We therefore list only the parameters of two observations. These data are clearly not sufficient to identify jet components confidently especially since the sum over the flux densities of the (unlabeled) components is different.\\
\begin{flushleft}
{\it 0954+658}\\
\end{flushleft}
Two identification scenarios resulting in either fast or slow jet component motion seem to be possible for this source. We consider the ``slow''-motion scenario to be more reliable.\\
\begin{flushleft}
{\it 1031+567}\\
\end{flushleft}
This source is a compact double. The core is in the middle of the source (see also Taylor et al. 1996).\\
\begin{flushleft}
{\it 1144+352}\\
\end{flushleft}
Although the source structure looks simple, the component identification in this source is complicated since the position angles of the individual components change and several unmatchable components appear in the second and third epoch. This nearby giant radio galaxy has been studied in detail by Giovannini et al. (1999).\\
\begin{flushleft}
{\it 1144+402}\\
\end{flushleft}
The jet components in this source are faint and the position angle changes make a reliable identification of a component difficult. It will not be considered any further.\\
\begin{flushleft}
{\it 1205+544}\\
\end{flushleft}
This is a convincing example for jet component separation on both sides of the reference point. No further information from the literature is available.\\
\begin{flushleft}
{\it 1206+415}\\
\end{flushleft}
We do not label any jet component in this source since the core separation of the jet component in the first epoch does not fit to the core separation of the jet components in the other two epochs.\\
\begin{flushleft}
{\it 1213+350}\\
\end{flushleft}
Jet components in this source are difficult to trace. We limit the identification to two jet components.\\
\begin{flushleft}
{\it 1246+586}\\
\end{flushleft}
We find two jet components approaching the reference point. No further information from the literature is available.\\
\begin{flushleft}
{\it 1413+373}\\
\end{flushleft}
It is difficult to trace jet components in the inner part ($< 5$ mas) of this source since some parameters of individual components are not consistent from epoch to epoch. We suspect a fast outflow in the inner part of this source or, alternatively backwards motion of several components.\\
\begin{flushleft}
{\it 1418+546}\\
\end{flushleft}
The whole source is fading in flux density, which complicates the jet component identification.\\
\begin{flushleft}
{\it 1442+637}\\
\end{flushleft}
This source is interesting with respect to the apparent velocities and directions of component motion. While C1 separates from the core, C2 remains at a fixed position, and components C3 and C4 approach the core. No further information on the pc-scale structure is available.\\
\begin{flushleft}
{\it 1459+480}\\
\end{flushleft}
Our results are in excellent agreement with those obtained at 8.55 GHz by FC97: components C1, C2, and C3 are at 1.1 mas, 3.3 mas, and 6.9 mas separation from the core and at 82$^{\circ}$, 86$^{\circ}$, and 71$^{\circ}$ position angle, respectively.\\
\begin{flushleft}
{\it 1531+772}\\
\end{flushleft}
The core is influenced by the jet component closest to the core, the flux densities of this component change significantly between the first and second epoch. We find no definable motion in this source.  \\
\begin{flushleft}
{\it 1543+517}\\
\end{flushleft}
Four of the six jet components apparently approach the reference point. Higher dynamic range observations to clarify the motion scenario in this source are clearly needed. No further information on the pc-scale structure in this source is available.\\
\begin{flushleft}
{\it 1624+416}\\
\end{flushleft}
The structure is complicated and aggravates the component identification.\\
\begin{flushleft}
{\it 1751+441}\\
\end{flushleft}
We trace the brightest jet component which most likely does not move at all. The other jet features are faint and ill defined in position.\\
\begin{flushleft}
{\it 1800+440}\\
\end{flushleft}
The epochs have such different resolutions that we cannot trace components reliably. Higher dynamic range observations are needed to enable a convincing identification of jet components.\\
\begin{flushleft}
{\it 1803+784}\\
\end{flushleft}
Although 1803+784 is a well-known source, it is still a matter of debate whether one or
more of its components are stationary (see Britzen et al. 2005a and 2005b and references therein; Kudryavtseva et al. 2006).\\
\begin{flushleft}
{\it 1839+389}\\
\end{flushleft}
The component is very faint and the tentative motion will not be taken into account in any further calculations.\\
\begin{flushleft}
{\it 1928+738}\\
\end{flushleft}
We find and trace six jet components in this prominent source, which is double-sided on kpc-scales (e.g., Hummel et al. 1992). C3 remains more or less at a similar core separation across the epochs. For this complex source structure higher dynamic range observations are needed to trace the jet components unambiguously.\\ 
\begin{flushleft}
{\it 1943+546}\\
\end{flushleft}
This source reveals a very complex but well characterizable pc-scale structure; we identify and trace confidently all jet components detected. This is a CSS source according to Saikia et al. (2001). \\
\begin{flushleft}
{\it 1946+708}\\
\end{flushleft}
This source might be the most complex source in this survey. We list the parameters for four epochs. It would be desirable to monitor this source with higher time sampling to trace all the components more confidently. This is a CSO and the true core is almost in the middle of the S-shaped symmetry (e.g., Taylor \& Vermeulen 1997; Peck \& Taylor 2001). We adopt the component definition from Taylor \& Vermeulen 1997.\\
\begin{flushleft}
{\it 1950+573}\\
\end{flushleft}
The data quality varies from epoch to epoch such that the identification of jet component presents problems. We assume that C3 in the third epoch is a blend of three jet components. 
\begin{flushleft}
{\it 2007+777}\\
\end{flushleft}
Severe flux-density changes of the reference component (r) are reported here. A reliable jet component identification is thus complicated.
\begin{flushleft}
{\it 2356+390}\\
\end{flushleft}
Only one jet component can reliably be traced. The remaining components are ill-defined.\\
\section{Conclusions}
We have been able to produce VLBI images of all but 5 of the 293
sources in the complete CJF sample. The selection of these bright,
flat-spectrum sources as targets for VLBI observations has obviously
been very successful ! All sources were observed at least three times.
In this paper we have discussed our observing and data analysis
procedures. We have presented all available images, as well as source
component parameters derived from model-fitting. We also present and
discuss the cross-epoch component identifications, which we have made in
preparation for a statistical analysis of component (superluminal)
motions. This will be presented in Paper~II. In
Paper~III we will present a correlation analysis
between the soft X-ray and VLBI properties of the CJF sources.
\begin{acknowledgements}
This work was supported by the European Commission, TMR Programme,
Research Network Contract ERBFMRXCT96-0034 ``CERES" and by the DLR, project 50QD0101.
S. Britzen acknowledges support by the Claussen-Simon-Stiftung.
Part of the data are based on observations with the 100-m telescope of the MPIfR (Max-Planck-Institut f\"ur Radioastronomie) at Effelsberg. This research has made use of the NASA/IPAC Extragalactic Database (NED) which 
is operated by the Jet Propulsion Laboratory, California Institute of Technology, 
under contract with the National Aeronautics and Space Administration. The National Radio Astronomy Observatory is operated by Associated Universities, Inc., under cooperative agreement with the National Science Foundation. The European VLBI Network is a joint facility of European, Chinese, South African and other radio astronomy institutes funded by their national research councils.
\end{acknowledgements}

\clearpage
\begin{figure*}[]
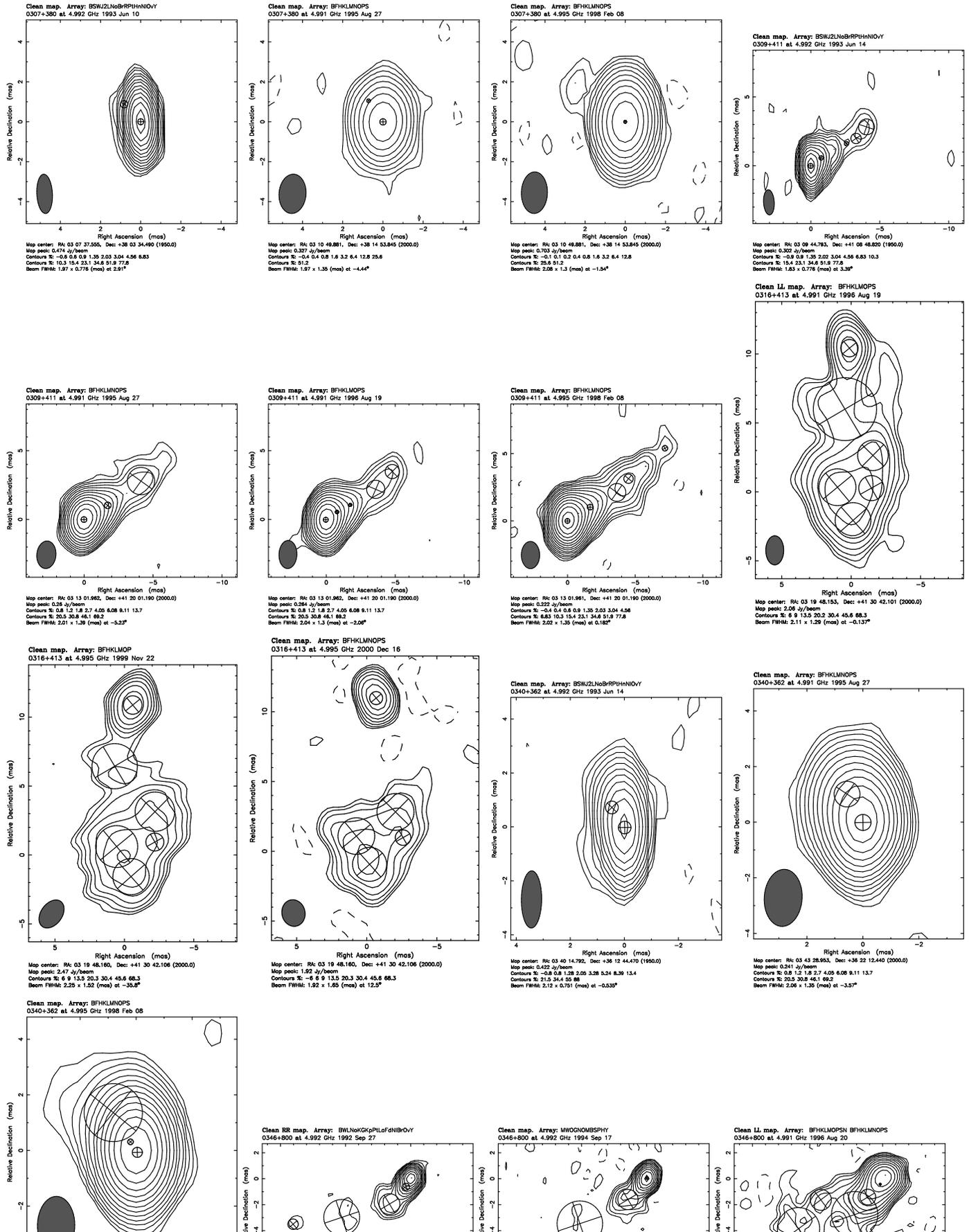

\caption{5 GHz VLBI images of CJF sources are shown, built up by restoring the model-fitted components, convolved with the clean beam, into the residual image, which was made by Fourier transforming the visibility data after first subtracting the model-fitted components in the uv-plane. Over-plotted we show symbols to represent the model components. We show the images of five sources in the printed version. The complete set of 894 images is available in electronic form at the CDS via anonymous ftp to cdsarc.u-strasbg.fr (130.79.128.5) or via http://cdsweb.u-strasbg.fr/cgi-bin/qcat?J/A+A/}
\label{maps}
\subfigure{\psfig{figure=2677f1.ps,width=4.45cm}}
\subfigure{\psfig{figure=2677f2.ps,width=4.45cm}}
\subfigure{\psfig{figure=2677f3.ps,width=4.45cm}}
\subfigure{\psfig{figure=2677f4.ps,width=4.45cm}}
\subfigure{\psfig{figure=2677f5.ps,width=4.45cm}}
\subfigure{\psfig{figure=2677f6.ps,width=4.45cm}}
\subfigure{\psfig{figure=2677f7.ps,width=4.45cm}}
\subfigure{\psfig{figure=2677f8.ps,width=4.45cm}}
\subfigure{\psfig{figure=2677f9.ps,width=4.45cm}}
\subfigure{\psfig{figure=2677f10.ps,width=4.45cm}}
\subfigure{\psfig{figure=2677f11.ps,width=4.45cm}}
\subfigure{\psfig{figure=2677f12.ps,width=4.45cm}}
\subfigure{\psfig{figure=2677f13.ps,width=4.45cm}}
\subfigure{\psfig{figure=2677f14.ps,width=4.45cm}}
\subfigure{\psfig{figure=2677f15.ps,width=4.45cm}}
\subfigure{\psfig{figure=2677f16.ps,width=4.45cm}}
\end{figure*}
\clearpage
\onecolumn
\tabcolsep0.3mm
\tablecaption{Results from model-fitting components in Fig. \ref{maps}. We only show the parameters of five sources in the printed version. The complete table for 288 sources is available in electronic form at the CDS via anonymous ftp to cdsarc.u-strasbg.fr (130.79.128.5) or via http://cdsweb.u-strasbg.fr/cgi-bin/qcat?J/A+A/} {\smallskip}
\centering
\tablehead{\noalign{\smallskip} \hline \noalign{\smallskip}
Source&Epoch&Id.&S&x&y&r&$\Theta$&M&$\chi^2$\\
\multicolumn{1}{c}{\small} & \multicolumn{1}{c}{\small} &
\multicolumn{1}{c}{\small} & 
\multicolumn{1}{r}{\tiny ~~~~~~~~~~~~~~[Jy]} & \multicolumn{1}{r}{\tiny [mas]} &
\multicolumn{1}{r}{\tiny [mas]} & \multicolumn{1}{r}{\tiny [mas]} &
\multicolumn{1}{r}{\tiny [deg]} & \multicolumn{1}{r}{\tiny [mas]}&\multicolumn{1}{l}{\small} \\
\noalign{\smallskip} \hline \hline  \noalign{\smallskip}}
\tabletail{\hline\multicolumn{8}{r}{\small continued on next page}\\}
\tablelasttail{\hline}
{\tiny
\begin{supertabular}{lllrrrrrrrr}
\label{modelfit}
0307+380 &93/06/10&r& 0.516$\pm$0.001&  0.00$\pm$0.00&  0.00$\pm$0.00& 0.00$\pm$0.00&   0.0$\pm$  0.0& 0.3$\pm$ 0.1&0.784 (9304)\\
JVAS J0310+3814&&C1& 0.006$\pm$0.001&  0.83$\pm$0.04&  0.88$\pm$0.10& 1.21$\pm$0.08&  43.3$\pm$  3.2& 0.4$\pm$ 0.2&\\
0307+380&95/08/27&r& 0.338$\pm$0.001&  0.00$\pm$0.00&  0.00$\pm$0.00& 0.00$\pm$0.00&   0.0$\pm$  0.0& 0.3$\pm$ 0.1&0.808 (7653)\\
&&C1& 0.005$\pm$0.001&  0.72$\pm$0.07&  1.06$\pm$0.09& 1.28$\pm$0.10&  34.2$\pm$  2.8& 0.2$\pm$ 0.2&\\
&&& 0.001$\pm$0.000&  3.01$\pm$0.16&  1.67$\pm$0.24& 3.44$\pm$0.17&  60.9$\pm$  3.8& 0.0$\pm$ 0.5&\\
0307+380&98/02/08&r& 0.710$\pm$0.000&  0.00$\pm$0.00&  0.00$\pm$0.00& 0.00$\pm$0.00&   0.0$\pm$  0.0& 0.2$\pm$ 0.1&0.765 (14336)\\
&&& 0.002$\pm$0.000&  2.50$\pm$0.05&  1.95$\pm$0.07& 3.17$\pm$0.05&  52.1$\pm$  1.2& 0.0$\pm$ 0.1&\\\\
0309+411 &93/06/14&r& 0.334$\pm$0.003&  0.00$\pm$0.00&  0.00$\pm$0.00& 0.00$\pm$0.00&   0.0$\pm$  0.0& 0.4$\pm$ 1.0&0.828 (7553)\\
JVAS J0313+4120&&C1& 0.082$\pm$0.002& -0.76$\pm$0.02&  0.58$\pm$0.02& 0.95$\pm$0.02& -52.7$\pm$  0.5& 0.3$\pm$ 0.1&\\
&&C2& 0.021$\pm$0.003& -1.67$\pm$0.11&  1.21$\pm$0.08& 2.07$\pm$0.13& -54.1$\pm$  1.0& 0.0$\pm$ 0.1&\\
&&C3& 0.005$\pm$0.009& -2.60$\pm$0.76&  1.61$\pm$0.55& 3.06$\pm$0.92& -58.1$\pm$  3.4& 0.3$\pm$ 0.7&\\
&&C4& 0.005$\pm$0.010& -3.29$\pm$0.98&  1.99$\pm$0.63& 3.85$\pm$1.12& -58.8$\pm$  4.7& 0.8$\pm$ 0.8&\\
&&C4& 0.013$\pm$0.005& -4.05$\pm$0.12&  2.83$\pm$0.18& 4.94$\pm$0.18& -55.0$\pm$  1.3& 1.2$\pm$ 0.3&\\
0309+411&95/08/27&r& 0.264$\pm$0.003&  0.00$\pm$0.00&  0.00$\pm$0.00& 0.00$\pm$0.00&   0.0$\pm$  0.0& 0.4$\pm$ 1.0&1.016 (8029)\\
&&C1& 0.039$\pm$0.006& -0.86$\pm$0.06&  0.58$\pm$0.04& 1.04$\pm$0.07& -56.3$\pm$  0.7& 0.0$\pm$ 0.1&\\
&&C2& 0.023$\pm$0.006& -1.73$\pm$0.20&  1.03$\pm$0.14& 2.01$\pm$0.24& -59.2$\pm$  1.2& 0.5$\pm$ 0.1&\\
&&C3& 0.001$\pm$0.012& -2.35$\pm$2.60&  1.66$\pm$1.99& 2.88$\pm$3.07& -54.7$\pm$ 22.7& 0.0$\pm$ 2.3&\\
&&C4& 0.003$\pm$0.009& -3.08$\pm$0.71&  1.91$\pm$0.33& 3.62$\pm$0.74& -58.2$\pm$  4.2& 0.0$\pm$ 0.8&\\
&&C4& 0.023$\pm$0.002& -4.14$\pm$0.07&  2.79$\pm$0.05& 4.99$\pm$0.08& -56.0$\pm$  0.5& 2.0$\pm$ 0.2&\\
0309+411&96/08/19&r& 0.261$\pm$0.003&  0.00$\pm$0.00&  0.00$\pm$0.00& 0.00$\pm$0.00&   0.0$\pm$  0.0& 0.4$\pm$ 1.0&1.073 (10061)\\
&&C1& 0.044$\pm$0.003& -0.81$\pm$0.06&  0.56$\pm$0.03& 0.99$\pm$0.06& -55.3$\pm$  0.6& 0.3$\pm$ 0.1&\\
&&C2& 0.020$\pm$0.004& -1.78$\pm$0.16&  1.08$\pm$0.10& 2.08$\pm$0.18& -58.7$\pm$  0.7& 0.2$\pm$ 0.1&\\
&&C3& 0.001$\pm$0.005& -2.51$\pm$1.40&  1.60$\pm$1.22& 2.98$\pm$1.80& -57.5$\pm$  8.5& 0.0$\pm$ 1.5&\\
&&C4& 0.013$\pm$0.003& -3.62$\pm$0.18&  2.21$\pm$0.11& 4.24$\pm$0.20& -58.6$\pm$  0.9& 1.3$\pm$ 0.3&\\
&&C5& 0.011$\pm$0.002& -4.82$\pm$0.06&  3.50$\pm$0.09& 5.96$\pm$0.09& -54.0$\pm$  0.6& 1.0$\pm$ 0.2&\\
0309+411&98/02/08&r& 0.221$\pm$0.001&  0.00$\pm$0.00&  0.00$\pm$0.00& 0.00$\pm$0.00&   0.0$\pm$  0.0& 0.4$\pm$ 1.0&0.956 (14555)\\
&&C1& 0.035$\pm$0.003& -0.80$\pm$0.04&  0.57$\pm$0.02& 0.98$\pm$0.04& -54.5$\pm$  0.5& 0.0$\pm$ 0.1&\\
&&C2& 0.016$\pm$0.005& -1.68$\pm$0.20&  1.03$\pm$0.13& 1.97$\pm$0.24& -58.6$\pm$  0.6& 0.4$\pm$ 0.1&\\
&&C3& 0.004$\pm$0.007& -2.27$\pm$0.32&  1.47$\pm$0.29& 2.71$\pm$0.42& -57.1$\pm$  2.0& 0.0$\pm$ 0.3&\\
&&C4& 0.010$\pm$0.002& -3.66$\pm$0.11&  2.12$\pm$0.08& 4.23$\pm$0.13& -59.9$\pm$  0.6& 1.3$\pm$ 0.2&\\
&&C5& 0.007$\pm$0.002& -4.52$\pm$0.05&  3.14$\pm$0.07& 5.50$\pm$0.07& -55.2$\pm$  0.5& 0.7$\pm$ 0.2&\\
&&C5& 0.005$\pm$0.001& -5.78$\pm$0.04&  4.06$\pm$0.05& 7.06$\pm$0.05& -54.9$\pm$  0.3& 0.0$\pm$ 0.1&\\
&&& 0.002$\pm$0.000& -7.21$\pm$0.08&  5.41$\pm$0.11& 9.01$\pm$0.10& -53.1$\pm$  0.6& 0.4$\pm$ 0.3&\\\\
0316+413 &96/08/19&r& 3.080$\pm$0.014&  0.00$\pm$0.00&  0.00$\pm$0.00& 0.00$\pm$0.00&   0.0$\pm$  0.0& 1.3$\pm$ 0.1&177.501 (10298)\\
JVAS J0319+4130&&C1& 4.462$\pm$0.035&  0.32$\pm$0.01& -4.42$\pm$0.01& 4.43$\pm$0.01& 175.8$\pm$  0.1& 4.6$\pm$ 0.1&\\
&&C2& 2.728$\pm$0.041& -1.69$\pm$0.01& -7.78$\pm$0.01& 7.96$\pm$0.01&-167.7$\pm$  0.1& 2.2$\pm$ 0.1&\\
&&C3& 6.270$\pm$0.095&  0.86$\pm$0.01&-10.09$\pm$0.01&10.12$\pm$0.01& 175.1$\pm$  0.1& 2.7$\pm$ 0.1&\\
&&C4& 2.023$\pm$0.057& -1.58$\pm$0.01&-10.15$\pm$0.02&10.27$\pm$0.02&-171.1$\pm$  0.1& 1.8$\pm$ 0.1&\\
&&C5& 4.302$\pm$0.070& -0.26$\pm$0.01&-12.49$\pm$0.01&12.50$\pm$0.01&-178.8$\pm$  0.0& 2.6$\pm$ 0.1&\\
0316+413&99/11/22&r& 3.680$\pm$0.012&  0.00$\pm$0.00&  0.00$\pm$0.00& 0.00$\pm$0.00&   0.0$\pm$  0.0& 1.3$\pm$ 0.1&214.447 (8772)\\
&&C1& 1.442$\pm$0.028&  1.30$\pm$0.02& -4.43$\pm$0.02& 4.62$\pm$0.01& 163.6$\pm$  0.3& 3.3$\pm$ 0.1&\\
&&C2& 2.408$\pm$0.055& -1.60$\pm$0.01& -7.74$\pm$0.02& 7.90$\pm$0.02&-168.3$\pm$  0.1& 2.9$\pm$ 0.1&\\
&&C3& 0.948$\pm$0.039& -1.59$\pm$0.01& -9.94$\pm$0.02&10.06$\pm$0.01&-170.9$\pm$  0.1& 1.3$\pm$ 0.1&\\
&&C4& 4.612$\pm$0.127&  1.16$\pm$0.02&-10.26$\pm$0.01&10.33$\pm$0.01& 173.6$\pm$  0.1& 3.0$\pm$ 0.1&\\
&&C5& 3.280$\pm$0.098&  0.12$\pm$0.01&-12.45$\pm$0.02&12.45$\pm$0.02& 179.4$\pm$  0.0& 2.6$\pm$ 0.1&\\
0316+413&2000/12/16&r& 1.977$\pm$0.010&  0.00$\pm$0.00&  0.00$\pm$0.00& 0.00$\pm$0.00&   0.0$\pm$  0.0& 0.9$\pm$ 0.1&607.242 (9240)\\
&&C2& 1.292$\pm$0.117& -1.38$\pm$0.04& -8.16$\pm$0.07& 8.28$\pm$0.07&-170.4$\pm$  0.3& 2.6$\pm$ 0.3&\\
&&C3& 1.403$\pm$0.189&  1.41$\pm$0.07& -9.89$\pm$0.06& 9.99$\pm$0.06& 171.9$\pm$  0.5& 2.6$\pm$ 0.3&\\
&&C4& 0.369$\pm$0.068& -1.92$\pm$0.03&-10.00$\pm$0.05&10.19$\pm$0.05&-169.1$\pm$  0.2& 1.1$\pm$ 0.2&\\
&&C5& 2.389$\pm$0.151&  0.47$\pm$0.02&-11.92$\pm$0.04&11.93$\pm$0.04& 177.8$\pm$  0.1& 2.4$\pm$ 0.1&\\\\
0340+362 &93/06/14&r& 0.475$\pm$0.002&  0.00$\pm$0.00&  0.00$\pm$0.00& 0.00$\pm$0.00&   0.0$\pm$  0.0& 0.5$\pm$ 0.1&0.875 (8100)\\
JVAS J0343+3622&&C1& 0.078$\pm$0.002&  0.47$\pm$0.01&  0.74$\pm$0.01& 0.88$\pm$0.01&  32.5$\pm$  0.4& 0.5$\pm$ 0.1&\\
0340+362&95/08/27&r& 0.251$\pm$0.001&  0.00$\pm$0.00&  0.00$\pm$0.00& 0.00$\pm$0.00&   0.0$\pm$  0.0& 0.6$\pm$ 0.1&1.019 (8772)\\
&&C1& 0.062$\pm$0.002&  0.57$\pm$0.01&  1.02$\pm$0.01& 1.17$\pm$0.01&  29.1$\pm$  0.4& 0.9$\pm$ 0.1&\\
0340+362&98/02/08&r& 0.280$\pm$0.014&  0.00$\pm$0.00&  0.00$\pm$0.01& 0.00$\pm$0.00&   0.0$\pm$  0.0& 0.4$\pm$ 0.1&0.883 (13198)\\
&&C1& 0.065$\pm$0.015&  0.24$\pm$0.02&  0.38$\pm$0.03& 0.45$\pm$0.03&  32.9$\pm$  0.9& 0.2$\pm$ 0.1&\\
&&C1& 0.025$\pm$0.001&  0.86$\pm$0.02&  1.43$\pm$0.03& 1.67$\pm$0.04&  31.0$\pm$  0.6& 2.1$\pm$ 0.1&\\\\
0346+800 &92/09/27&r& 0.235$\pm$0.002&  0.00$\pm$0.00&  0.00$\pm$0.00& 0.00$\pm$0.00&   0.0$\pm$  0.0& 0.0$\pm$ 0.1&0.903 (9606)\\
JVAS J0354+8009&&C1& 0.074$\pm$0.002&  0.40$\pm$0.00& -0.68$\pm$0.01& 0.79$\pm$0.01& 149.6$\pm$  0.2& 0.5$\pm$ 0.1&\\
&&C2& 0.055$\pm$0.002&  1.67$\pm$0.03& -2.01$\pm$0.03& 2.62$\pm$0.03& 140.3$\pm$  0.5& 1.6$\pm$ 0.1&\\
&&& 0.021$\pm$0.002&  5.34$\pm$0.10& -3.04$\pm$0.09& 6.14$\pm$0.10& 119.7$\pm$  0.9& 2.8$\pm$ 0.9&\\
&&& 0.005$\pm$0.001&  9.08$\pm$0.07& -3.56$\pm$0.08& 9.75$\pm$0.07& 111.4$\pm$  0.4& 0.8$\pm$ 0.1&\\
0346+800&94/09/17&r& 0.219$\pm$0.004&  0.00$\pm$0.00&  0.00$\pm$0.01& 0.00$\pm$0.00&   0.0$\pm$  0.0& 0.2$\pm$ 0.1&0.998 (4204)\\
&&C1& 0.034$\pm$0.004&  0.52$\pm$0.03& -0.73$\pm$0.03& 0.89$\pm$0.03& 144.4$\pm$  1.2& 0.0$\pm$ 0.1&\\
&&C2& 0.053$\pm$0.004&  1.47$\pm$0.05& -1.65$\pm$0.05& 2.21$\pm$0.06& 138.3$\pm$  1.1& 1.6$\pm$ 0.2&\\
&&& 0.017$\pm$0.005&  5.17$\pm$0.42& -3.51$\pm$0.41& 6.25$\pm$0.45& 124.2$\pm$  3.3& 3.5$\pm$ 4.6&\\
0346+800&96/08/20&r& 0.076$\pm$0.004&  0.00$\pm$0.01&  0.00$\pm$0.01& 0.00$\pm$0.00&   0.0$\pm$  0.0& 0.0$\pm$ 0.1&1.064 (13564)\\
&&C1& 0.041$\pm$0.005&  0.37$\pm$0.02& -0.70$\pm$0.03& 0.79$\pm$0.02& 152.3$\pm$  0.8& 0.1$\pm$ 0.1&\\
&&C2& 0.043$\pm$0.003&  1.32$\pm$0.02& -1.70$\pm$0.02& 2.15$\pm$0.03& 142.1$\pm$  0.4& 1.2$\pm$ 0.1&\\
&&C2& 0.019$\pm$0.005&  2.33$\pm$0.21& -3.05$\pm$0.27& 3.83$\pm$0.25& 142.6$\pm$  3.4& 3.6$\pm$ 1.4&\\
&&& 0.007$\pm$0.002&  4.93$\pm$0.13& -1.91$\pm$0.09& 5.29$\pm$0.12& 111.1$\pm$  1.0& 1.6$\pm$ 0.6&\\
&&& 0.014$\pm$0.002&  6.64$\pm$0.18& -4.13$\pm$0.14& 7.82$\pm$0.20& 121.9$\pm$  0.8& 3.3$\pm$ 1.2&\\
\noalign{\smallskip}
\hline
\end{supertabular}
}
\vspace*{1.5cm}
\end{document}